\documentclass[sigconf, nonacm]{acmart}

\usepackage{algorithm}
\usepackage[noEnd=false,indLines=false]{algpseudocodex}
\usepackage{subcaption}
\usepackage{booktabs}
\usepackage{multirow}
\usepackage{mathtools}
\usepackage{xcolor}
\usepackage{xspace}
\usepackage{enumitem}

\graphicspath{{./}{figures/}}

\newcommand\vldbdoi{XX.XX/XXX.XX}
\newcommand\vldbpages{XXX-XXX}
\newcommand\vldbvolume{XX}
\newcommand\vldbissue{X}
\newcommand\vldbyear{20XX}
\newcommand\vldbauthors{\authors}
\newcommand\vldbtitle{\shorttitle}
\newcommand\vldbavailabilityurl{}
\newcommand\vldbpagestyle{plain}

\newcommand{\merit}{M\textsc{erit}\xspace}

\newcommand{\resultsmallpanela}[2]{%
  \begin{subfigure}[t]{0.48\columnwidth}
    \centering
    \includegraphics[width=\linewidth]{figures/experiment_results/#1}
    \captionsetup{skip=1ex,justification=centering,margin={0pt,15pt}}
    \caption{#2}
  \end{subfigure}%
}
\newcommand{\resultsixpanel}[2]{%
  \begin{subfigure}[t]{0.165\textwidth}
    \centering
    \includegraphics[width=\linewidth]{figures/experiment_results/#1}
    \captionsetup{font=scriptsize,skip=1ex,justification=centering,singlelinecheck=false, margin={0pt,20pt}}
    \caption{#2}
  \end{subfigure}%
}
\newcommand{\resultsixpanela}[2]{%
  \begin{subfigure}[t]{0.165\textwidth}
    \centering
    \includegraphics[width=\linewidth]{figures/experiment_results/#1}
    \captionsetup{font=scriptsize,skip=1ex,justification=centering,singlelinecheck=false, margin={20pt,0pt}}
    \caption{#2}
  \end{subfigure}%
}
\newcommand{\figureTopMargin}{\vspace{-1ex}}
\newcommand{\figureCaptionMargin}{\vspace{-1ex}}
\newcommand{\figureBelowMargin}{\vspace{-2ex}}

\newcommand{\tableBelowMargin}{\vspace{-3ex}}

\algrenewcommand\algorithmiccomment[1]{\hfill\textit{$\triangleright$ #1}}

\begin{document}

\title[MERIT: Efficient In-Place Deletion for Dynamic Graph-Based Approximate Nearest Neighbor Indexes]{MERIT: Efficient In-Place Deletion for Dynamic \\ Graph-Based Approximate Nearest Neighbor Indexes}

\author{Zekai Wu}
\affiliation{%
  \institution{Beijing Institute of Technology}
  \institution{Tongji University}
}
\email{zekaiwu0629@gmail.com}

\author{Jiabao Jin}
\affiliation{
  \institution{Tongji University}
}
\email{jiabaojin0723@gmail.com}

\author{Peng Cheng}
\authornote{Corresponding Author.}
\affiliation{
  \institution{Tongji University}
}
\email{cspcheng@tongji.edu.cn}

\author{Wangze Ni}
\affiliation{
  \institution{Zhejiang University}
}
\email{niwangze@zju.edu.cn}

\author{Haoyang Li}
\affiliation{
  \institution{Hong Kong Polytechnic University}
}
\email{haoyang-comp.li@polyu.edu.hk}

\author{Lei Chen}
\affiliation{
  \institution{Hong Kong University of Science and Technology}
}
\email{leichen@cse.ust.hk}

\author{Junjie Yao}
\affiliation{%
  \institution{East China Normal University}
}
\email{junjie.yao@sei.ecnu.edu.cn}

\author{Jingkuan Song}
\affiliation{
  \institution{Tongji University}
}
\email{jingkuan.song@gmail.com}

\author{Heng Tao Shen}
\affiliation{
  \institution{Tongji University}
}
\email{shenhengtao@hotmail.com}

\begin{abstract}
  Graph-based indexes have become the dominant approach to approximate nearest neighbor search (ANNS) over high-dimensional data and play a crucial role in real-world applications such as retrieval-augmented generation, recommendation systems, and vector databases. Despite extensive progress in static graph construction and search, efficient in-place deletion remains challenging because obsolete vectors must be removed without allowing stale incoming edges to consume search capacity or expensive graph-wide maintenance to interrupt online services, e.g., retrieval-augmented generation (RAG) and recommendation platforms. To address this problem, we propose \merit (\textbf{M}ST-based \textbf{E}fficient \textbf{R}epair with \textbf{I}n-place upda\textbf{T}es), an in-place update framework with three core techniques: (1) bounded search-based recovery that combines a deleted vertex's outgoing neighbors with its readily searchable in-neighbors, (2) $k_r$-Minimum Spanning Tree (MST) local repair that promotes local connectivity while retaining multiple routing choices for graph search, and (3) versioned-edge invalidation that immediately filters all stale incoming edges to the deleted vertex and progressively removes them as adjacency lists are rewritten. Its integration with the hierarchical HNSW index and the single-layer Vamana index demonstrates applicability across distinct graph structures. Extensive experiments on multiple real-world datasets show that \merit processes deletion at nearly the cost of inserting one vector, achieves up to $3.02\times$--$18.87\times$ faster deletion than state-of-the-art (SOTA) methods, and keeps search recall stable or even improves it as deletions accumulate.
\end{abstract}

\maketitle

\pagestyle{\vldbpagestyle}
\begingroup\small\noindent\raggedright\textbf{PVLDB Reference Format:}\\
\vldbauthors. \vldbtitle. PVLDB, \vldbvolume(\vldbissue): \vldbpages, \vldbyear.\\
\href{https://doi.org/\vldbdoi}{doi:\vldbdoi}
\endgroup
\begingroup
\renewcommand\thefootnote{}\footnote{\noindent
  This work is licensed under the Creative Commons BY-NC-ND 4.0 International License. Visit \url{https://creativecommons.org/licenses/by-nc-nd/4.0/} to view a copy of this license. For any use beyond those covered by this license, obtain permission by emailing \href{mailto:info@vldb.org}{info@vldb.org}. Copyright is held by the owner/author(s). Publication rights licensed to the VLDB Endowment.\\
  \raggedright Proceedings of the VLDB Endowment, Vol.~\vldbvolume, No.~\vldbissue\ ISSN 2150-8097.\\
  \href{https://doi.org/\vldbdoi}{doi:\vldbdoi}\\
}\addtocounter{footnote}{-1}\endgroup

\ifdefempty{\vldbavailabilityurl}{}{%
  \vspace{.3cm}%
  \begingroup\small\noindent\raggedright\textbf{PVLDB Artifact Availability:}\\
  The source code, data, and/or other artifacts have been made available at \url{\vldbavailabilityurl}.%
  \endgroup}

\section{Introduction}

Vector representations have become the common substrate for modern data-intensive applications. Text passages, web pages, images, audio segments, videos, and other multimodal objects are routinely embedded into high-dimensional spaces where geometric proximity captures semantic similarity. As a result, approximate nearest neighbor search (ANNS)~\cite{arya1993approximate} is no longer a specialized numerical primitive, but a core operator in machine learning~\cite{cost1993weighted, wang2020survey}, information retrieval~\cite{wang2012scalable, zhu2019accelerating}, recommendation systems~\cite{das2007google,meng2020pmd}, vector databases~\cite{wang2021milvus}, and retrieval-augmented generation (RAG)~\cite{dobson2024scaling, asai2023retrieval}.

Among the major ANNS index families, graph-based methods~\cite{malkov2018efficient, fu2017fast, jayaram2019diskann, peng2023efficient,yu2025approximate} have become the dominant choice in practice~\cite{wang2021milvus,zhong2025vsag}. Their appeal is empirical but robust. Sparse proximity graphs provide short navigable paths through high-dimensional data, enabling beam-search-like traversal to achieve high recall with low latency and high throughput. Compared with tree~\cite{bentley1975multidimensional,silpa2008optimised,muja2014scalable}, hashing~\cite{indyk1998approximate,sun2014srs,huang2015query}, and inverted index-based methods~\cite{babenko2014inverted,babenko2016efficient,li2025sindi}, graph-based indexes often provide the best operating point when an online service must trade search quality against tail latency and memory footprint, according to recent benchmarks~\cite{wang2021comprehensive, aumuller2020ann}.

\begin{figure}[t]
  \centering
  \figureTopMargin
  \includegraphics[width=\columnwidth]{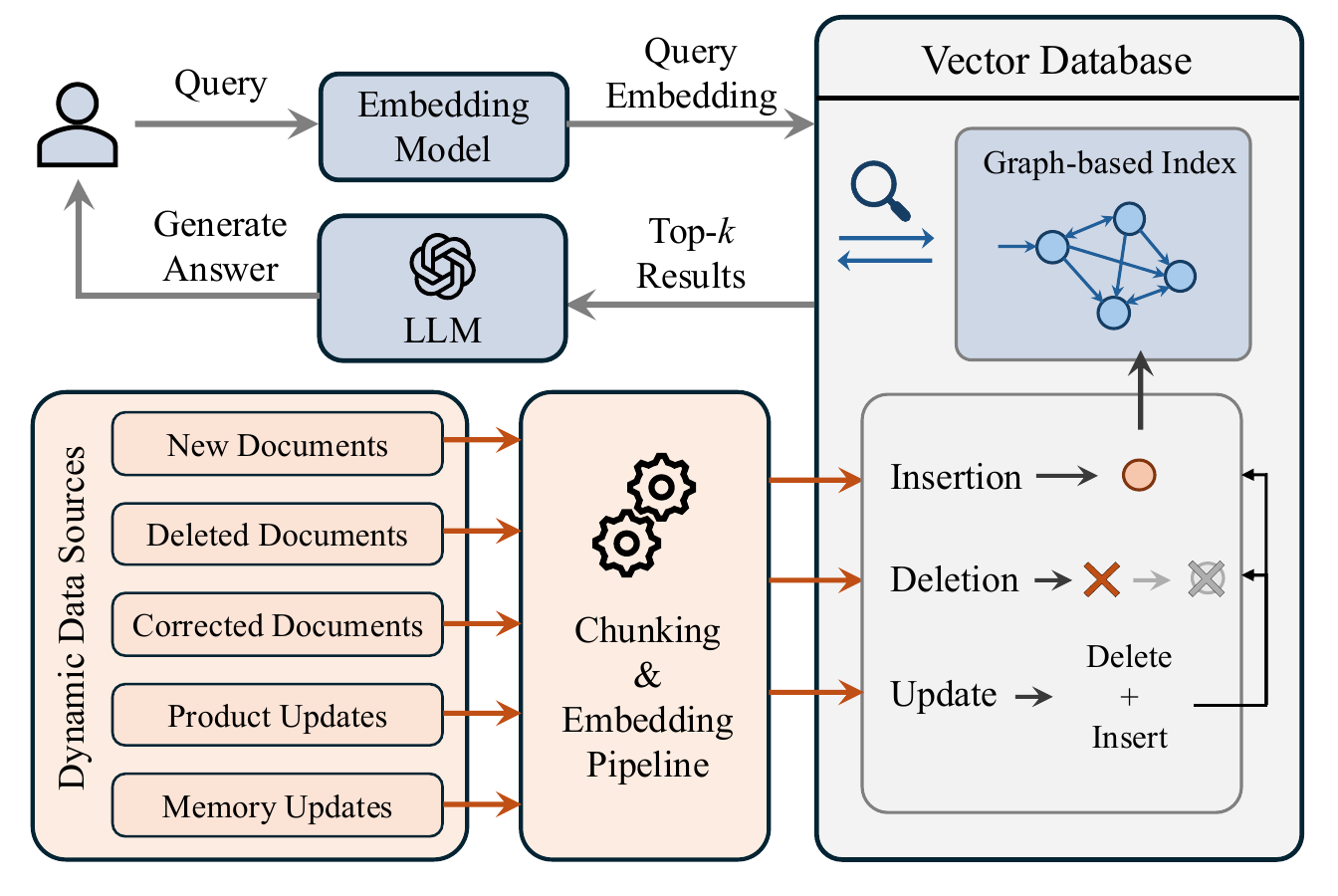}
  \figureCaptionMargin\vspace{-4ex}
  \caption{Dynamic RAG workloads.}
  \vspace{-4ex}
  \label{fig:intro}
\end{figure}

The success of graph-based ANNS has been established mostly under a static index assumption that a graph is built once and serves a large number of queries. However, recent vector-search deployments increasingly violate this assumption. As shown in Figure~\ref{fig:intro}, RAG knowledge bases need to be refreshed as documents are created, corrected, or removed; search and recommendation catalogs change continuously; user-specific memories are rewritten as users interact with the system; some privacy and compliance workflows require physically removing expired vectors~\cite{yang2024crag, zhao2026retrieval, cheng2025survey}. These workloads call for a dynamic index that supports efficient insertions and deletions while simultaneously serving online queries.

Insertion is relatively well aligned with graph-index design~\cite{malkov2018efficient,fu2017fast,jayaram2019diskann,peng2023efficient,fu2021high,malkov2014approximate}. A new vector can be inserted into the existing graph through searching, then connected to a small set of candidate neighbors, and incorporated by local edge updates. Deletion is substantially harder. Removing a vertex not only removes the vector from the searchable set, but also breaks the graph structure that used the vertex as a routing point. If the system only marks the vertex as deleted, queries may still waste candidate slots and distance computations on stale vertices. If the system repairs every affected edge, deletion can become much more expensive than insertion, creating update stalls and service-level instability. Vector replacement inherits the same difficulty because it is commonly implemented as a deletion followed by an insertion.

Existing production systems therefore rely on compromises~\cite{wang2021milvus,douze2024faiss, ozturk2024performance}. Soft deletion~\cite{malkov2018efficient} keeps updates cheap by leaving the deleted vertex in the graph, but stale vertices accumulate and progressively degrade recall and search efficiency. Segmented or log-structured indexing~\cite{o1996log, wang2021milvus} isolates updates in fresh segments, but query processing must merge results across segments, and obsolete segments eventually require consolidation. Periodic rebuilding~\cite{singh2021freshdiskann} restores graph quality, but it consumes additional resources and can introduce latency spikes. Recent in-place update schemes~\cite{xu2025place} reduce the need for rebuilding, yet deletion remains the bottleneck because identifying and repairing all affected edges is still expensive.

\textbf{The root cause is the asymmetric representation used by graph ANN indexes.} For compactness and update locality, each vertex stores only its outgoing neighbors. The incoming neighbors of a vertex are not stored explicitly. This design is efficient for search and insertion, but it makes deletion fundamentally indirect. After deleting a vertex $v_d$, the index does not know which other vertices contain outgoing edges to $v_d$. These vertices are precisely the ones whose adjacency lists must be cleaned or repaired.

Given this asymmetry, deletion has only two direct ways to find the affected vertices. Firstly, the index can store explicitly reverse adjacency, which doubles structural metadata, complicates memory layout, and introduces additional write contention because every insertion, deletion, and neighbor pruning decision must update two coupled adjacency views. Alternatively, the index can recover reverse neighbors by scanning the graph or issuing extra searches. Although this avoids explicit reverse edges, it turns deletion into an expensive maintenance task that must locate the affected vertices and then update and prune their adjacency lists one by one.

\textbf{This paper studies how to make deletions immediately visible to queries, maintain long-term recall stability, and achieve deletion efficiency comparable to insertion.} We present \merit, an in-place update mechanism for dynamic graph-based ANNS indexes that sustains efficient updates over long-running workloads. \merit decomposes deletion into three ordered stages: (1) marking the target vertex invalid to immediately remove it from query results while snapshotting its outgoing neighbors as repair seeds; (2) performing bounded searches to identify additional affected vertices and synchronously restoring their local topology through a $k$-MST construction; and (3) incrementing the deleted vertex's version to invalidate all residual stale incoming edges. By repairing a subset of the surviving outgoing and searchable incoming neighborhood, \merit preserves effective navigation paths without requiring exhaustive reverse-edge discovery, while versioned edges safely invalidate stale incoming edges missed during recovery. This design enables \merit to achieve robust long-term recall stability with deletion efficiency comparable to insertion. This paper makes the following contributions.

\begin{enumerate}[leftmargin=1.2em]
  \item We first analyze the in-place deletion problem for dynamic graph-based ANNS in \S\ref{sec:preliminaries}. Then, we present an experimental study of sustained updates in \S\ref{sec:challenges} to analyze why existing SOTA approaches suffer from unstable recall and disproportionately high deletion latency.

  \item In \S\ref{sec:merit}, we propose \merit, an in-place update mechanism that supports efficient deletion. \merit makes deleted vertices immediately invisible to search and preserves routing topology under bounded repair budget without explicitly maintaining incoming-neighbor lists.

  \item In \S\ref{sec:analysis} we analyze the complexity and long-term stability of \merit. In \S\ref{sec:experiments}, extensive experiments against SOTA dynamic graph ANNS baselines show that \merit substantially reduces deletion latency while maintaining stable high recall under sustained churn over several public datasets.
\end{enumerate}

The remainder of the paper is organized as follows. \S\ref{sec:related} surveys related work and positions our contribution. \S\ref{sec:conclusion} concludes the paper.

\section{Background and Problem Formulation}\label{sec:preliminaries}

This section first introduces the background on graph-based ANNS and then formulates the dynamic update model. Table~\ref{tab:symbols} summarizes the key symbols used in the paper and their descriptions.

\begin{table}[t]
  \vspace{-1ex}
  \centering
  {\small
    \caption{\small Symbols and Descriptions}
    \vspace{-2ex}
    \label{tab:symbols}
    \vspace{-1ex}
    \begin{tabular}{l|p{0.66\columnwidth}}
      \toprule
      {\bf Symbol}          & \multicolumn{1}{c}{\bf Description}                                 \\
      \midrule
      $\mathcal{X}$         & the base vector dataset                                             \\
      $\vec{x}, \vec{x}_q$  & a data vector and a query vector                                    \\
      $G=(V,E)$             & a graph with vertex set $V$ and edge set $E$                        \\
      $n=|V|$               & the number of vertices in the graph                                 \\
      $N_{\mathrm{}}(u)$    & the outgoing neighbors of vertex $u$                                \\
      $I(u)$                & the incoming-neighbor set of $u$                                    \\
      $\delta(\cdot,\cdot)$ & the distance function in the vector space                           \\
      $L$                   & the candidate pool size in graph search                             \\
      $M$                   & the maximum out-degree of the graph                                 \\
      $v_d$                 & the vertex to be deleted                                            \\
      $C$                   & the local repair candidate set                                      \\
      $\widehat{I}(v_d)$    & the recovered approximate in-neighbors of $v_d$                     \\
      $ef'_c$               & the beam width used for repair candidate discovery                  \\
      $k_r$                 & the number of repair edges retained per candidate in $k$-MST repair \\
      $H$                   & the repaired local subgraph induced on $C$                          \\
      \bottomrule
    \end{tabular}
  }
  \vspace{-1ex}
\end{table}

\subsection{Approximate Nearest Neighbor Search}

\subsubsection{ANNS Definition.}\label{subsubsec:anns}

Let $d \in \mathbb{N}$ denote the vector dimension and let $\mathcal{X}=\{\vec{x}_1,\vec{x}_2,\ldots,\vec{x}_n\}\subset\mathbb{R}^d$ be a finite dataset of $n$ vectors. For any two vectors $\vec{x}_a,\vec{x}_b\in\mathbb{R}^d$, let $\delta(\vec{x}_a,\vec{x}_b)$ denote their distance under a fixed metric or similarity induced distance, such as Euclidean distance or cosine distance. Given a query vector $\vec{x}_q\in\mathbb{R}^d$ and an integer $k\le n$, exact $k$ nearest neighbor search returns the set $\mathcal{R}_k(\vec{x}_q)\subseteq\mathcal{X}$ of size $k$ such that every returned vector is no farther from $\vec{x}_q$ than any nonreturned vector.

\begin{definition}[$k$ nearest neighbor search]
  \label{def:knns}
  Given $\mathcal{X}$, $\vec{x}_q$, $k$, and $\delta$, the exact $k$ nearest neighbor result is
  \vspace{-1ex}
  \begin{equation}
    \mathcal{R}_k(\vec{x}_q)
    = \arg\min_{\substack{\mathcal{R}\subseteq\mathcal{X}\\|\mathcal{R}|=k}}
    \sum_{\vec{x}\in\mathcal{R}}\delta(\vec{x}_q,\vec{x}).
  \end{equation}
  Equivalently, for any $\vec{x}_r\in\mathcal{R}_k(\vec{x}_q)$ and any $\vec{x}_s\in\mathcal{X}\setminus\mathcal{R}_k(\vec{x}_q)$, we have $\delta(\vec{x}_q,\vec{x}_r)\le\delta(\vec{x}_q,\vec{x}_s)$.
\end{definition}

In modern vector search workloads, both $n$ and $d$ can be large enough that exact $k$ nearest neighbor search is prohibitively expensive~\cite{prokhorenkova2020graph,indyk1998approximate}. ANNS therefore builds an index over $\mathcal{X}$ and uses the index to avoid exhaustive distance evaluation, trading exactness for lower latency, higher throughput, or fewer distance computations.

\begin{definition}[$k$ approximate nearest neighbor search]
  \label{def:anns}
  Given $\mathcal{X}$, $\vec{x}_q$, $k$, and $\delta$, ANNS visits a candidate subset $\mathcal{C}\subseteq\mathcal{X}$ instead of exhaustively scanning all vectors. It returns an approximate result $\widehat{\mathcal{R}}_k(\vec{x}_q)\subseteq\mathcal{C}$ with $|\widehat{\mathcal{R}}_k(\vec{x}_q)|=k$, usually by ranking the visited candidates according to $\delta(\vec{x}_q,\vec{x})$.
\end{definition}

We use two query-side metrics throughout the paper.
\begin{enumerate}[leftmargin=1.2em]
  \item \textbf{Recall.}
        Recall measures how much of the exact top-$k$ neighborhood is recovered by the approximate search. For $k$ nearest neighbor search, we use $\mathrm{Recall@}k$, defined as
        \vspace{-1ex}
        \begin{equation}
          \mathrm{Recall@}k =
          \frac{|\widehat{\mathcal{R}}_k(\vec{x}_q)\cap\mathcal{R}_k(\vec{x}_q)|}{k},
        \end{equation}
        where $\widehat{\mathcal{R}}_k(\vec{x}_q)$ is the approximate result set and $\mathcal{R}_k(\vec{x}_q)$ is the exact ground-truth result set, with $|\widehat{\mathcal{R}}_k(\vec{x}_q)|=|\mathcal{R}_k(\vec{x}_q)|=k$.

  \item \textbf{Query efficiency.}
        We report queries per second (QPS) for end-to-end search throughput and the number of distance computations (NDC) for algorithmic search work. NDC captures how many candidate vectors the graph traversal evaluates before returning the result.
\end{enumerate}

\subsubsection{Graph-Based ANNS.}

Graph-based indexes are among the most effective ANNS structures in practice~\cite{aumuller2020ann,li2019approximate,wang2021comprehensive}. They represent the dataset $\mathcal{X}$ as a sparse proximity graph $G=(V,E)$, where each vertex $v_i\in V$ corresponds to a vector $\vec{x}_i\in\mathcal{X}$ and each edge encodes a local proximity relation under $\delta$. In practical indexes, $G$ is stored as a directed graph. Each vertex $u$ maintains a bounded outgoing neighbor set $N(u)=\{v\mid (u,v)\in E\}$, and search traverses these adjacency lists from one or more entry points. A high-quality graph must balance two structural goals. Local edges should connect nearby vectors so that the final candidates are accurate, while navigable shortcuts should provide short paths across the graph and prevent the traversal from being trapped in a small region.

Most graph ANNS methods~\cite{fu2017fast,malkov2018efficient,jayaram2019diskann,peng2023efficient} follow a common construction pattern. For each vertex, the index first obtains a candidate neighbor set through search, incremental insertion, or refinement of an approximate $k$ nearest neighbor graph, and then applies a neighbor selection rule to keep only a small number of informative edges~\cite{wang2021comprehensive,yang2024revisiting}. HNSW~\cite{malkov2018efficient} builds a hierarchy of proximity graphs whose upper layers provide routing and whose base layer supports accurate search. Vamana~\cite{jayaram2019diskann} starts from a coarse graph and uses robust pruning to obtain a bounded-degree graph for large-scale search. NSG~\cite{fu2017fast} refines an approximate $k$ nearest neighbor graph ($k$-NNG) with pruning rules inspired by navigability. These pruning strategies include relative neighborhood graph (RNG) or monotonic relative neighborhood graph (MRNG) criteria~\cite{toussaint1980relative,fu2017fast}. Despite their algorithmic differences, these indexing approaches converge on a common representation in which outgoing adjacency lists are explicitly maintained, while incoming edges are not stored.

\subsubsection{Search Algorithm.}

Graph-based ANNS is typically executed by a greedy best-first traversal strategy known as beam search~\cite{yang2024revisiting}. Starting from an entry point $ep$, beam search maintains a bounded candidate set of size $L$, also called the beam width, and repeatedly explores promising vertices in distance order. Its effectiveness depends on short paths and long-range shortcuts in the graph. When these paths are preserved, beam search can reach high-recall answers with relatively few distance evaluations~\cite{malkov2018efficient,peng2023efficient}.

\begin{algorithm}[t]
  \small
  \caption{\textsc{KnnSearch}($\vec{x}_q, G, L, k, ep$)}
  \label{alg:anns}
  \begin{algorithmic}[1]
    \Require query vector $\vec{x}_q$, graph index $G=(V,E)$, search pool size $L\ge k$, entry point $ep$
    \Ensure approximate $k$ nearest neighbors of $\vec{x}_q$
    \State $C \gets \{ep\}$; mark $ep$ as visited
    \State $i \gets 0$
    \While{$i < |C|$ \textbf{and} $i < L$}
    \State $u \gets C[i]$
    \State mark $u$ as expanded
    \ForAll{$v \in N(u)$ \textbf{and} $v$ is not visited}
    \State insert $v$ into $C$ with distance $\delta(\vec{x}_q,\vec{x}_v)$
    \EndFor
    \State sort $C$ by increasing distance to $\vec{x}_q$ and keep the top-$L$ candidates
    \State $i \gets$ the index of the first unexpanded vertex in $C$
    \EndWhile
    \State \Return the first $k$ vertices in $C$
  \end{algorithmic}
\end{algorithm}

Algorithm~\ref{alg:anns} summarizes the canonical search routine. It initializes the candidate set with the entry point and marks it as visited (Line~1), then sets the scan pointer to the first candidate (Line~2). The main loop expands the closest unexpanded vertex in the current candidate set (Lines~3--5), evaluates its unvisited outgoing neighbors and inserts them into the candidate set (Lines~6--8), and keeps only the $L$ closest candidates before advancing to the next unexpanded vertex (Lines~9--10). The algorithm returns the first $k$ candidates as the approximate result (Line~12). In the rest of this paper, we use \textsc{KnnSearch} to refer to this search pattern when discussing queries, insertions, and repairs.

The algorithm also exposes why deletions are harmful. A stale outgoing edge can still be read during neighbor expansion (Lines~6--7), and a deleted vertex can still occupy the candidate set after insertion and pruning (Lines~7--9) unless it is filtered. Even when filtering prevents invalid results, the search budget spent on stale vertices is lost. Under a fixed beam width, this reduces the probability that the search reaches the true nearest-neighbor region.

\subsection{Update Model}\label{subsec:update-model}

Classical ANNS indexes are often described under a static setting, where the dataset is fixed after index construction and the index only needs to answer queries. Modern vector-search deployments, especially RAG systems over continuously refreshed knowledge bases, operate on dynamic datasets. New objects may be inserted as documents arrive, obsolete or expired objects must be deleted, and existing objects may be replaced when their contents or embeddings change. Therefore, a dynamic ANNS index must support querying, insertion, and deletion. Let $G=(V,E)$ be a graph ANN index over $\mathcal{X}\subseteq\mathbb{R}^d$. We consider three online update operations:

\begin{itemize}[leftmargin=1.2em]
  \item \emph{Insert}$(\vec{x})$ adds a new vector to the indexed set by searching the existing graph for candidate neighbors, selecting a bounded outgoing neighborhood for the new vertex, and applying local edge updates to preserve navigability.

  \item \emph{Delete}$(v_d)$ removes an existing vertex from the logical dataset and repairs the graph so that subsequent queries neither return the deleted object nor rely on it as a persistent routing shortcut.

  \item \emph{Replace}$(v,\vec{x}')$ updates the vector value associated with an object. In most graph ANNS systems, this operation is implemented as \emph{Delete}$(v)$ followed by \emph{Insert}$(\vec{x}')$.
\end{itemize}

\begin{figure}[t]
  \centering
  \figureTopMargin
  \includegraphics[width=0.5\textwidth]{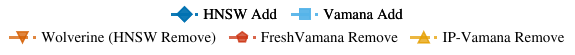}\\
  \begin{subfigure}{0.22\textwidth}
    \centering
    \includegraphics[width=\linewidth]{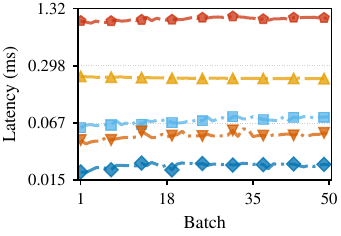}
    \caption{Deep10M, 30\% Deletion}
  \end{subfigure}
  \hfill
  \begin{subfigure}{0.22\textwidth}
    \centering
    \includegraphics[width=\linewidth]{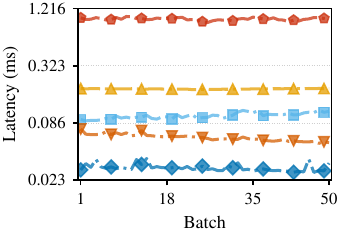}
    \caption{MSong, 30\% Deletion}
  \end{subfigure}
  \figureCaptionMargin\vspace{-2ex}
  \caption{Average insertion and deletion latency over update batches under the sliding-window workload.}
  \label{fig:update-latency-motivation}
  \figureBelowMargin
  \vspace{-2ex}
\end{figure}

An \emph{in-place deletion} immediately removes the deleted vertex $v_d$ from the current graph $G$ during the update stream, rather than deferring removal to a later rebuild or compaction phase. It must avoid the high storage and maintenance overhead of incoming edges, while repairing the local connectivity around $v_d$ so that the remaining graph continues to support accurate and efficient search.

\textbf{Update Efficiency.} We consider a batch of $b$ replacements, each implemented as one deletion followed by one insertion. Let $T_{\mathrm{ins}}$ and $T_{\mathrm{del}}$ denote the total elapsed times of the resulting $b$ insertions and $b$ deletions, respectively. We report the average insertion, deletion, and update latencies as
\[
  \begin{aligned}
    L_{\mathrm{ins}} & =T_{\mathrm{ins}}/b,                &
    L_{\mathrm{del}} & =T_{\mathrm{del}}/b,                &
    L_{\mathrm{upd}} & =L_{\mathrm{ins}}+L_{\mathrm{del}}.
  \end{aligned}\vspace{-1ex}
\]
Here, $L_{\mathrm{upd}}$ measures the maintenance latency of one delete--insert replacement to refresh one indexed vector in the dynamic workload.

\subsection{The Inefficiency of Current Approaches}\label{subsec:update-motivation}

Dynamic graph-based ANNS systems typically prioritize efficient insertion and query processing, while postponing the more complex deletion path through auxiliary mechanisms such as lazy deletion~\cite{singh2021freshdiskann}. This design introduces a pronounced asymmetry under sustained update workloads. Insertion latency remains close to the cost of one graph search followed by bounded neighbor updates. By contrast, deletion latency either becomes highly variable because the actual maintenance work is deferred as a batch job~\cite{liu2025wolverine}, or increases substantially when the system performs reverse-edge discovery and local graph repair during deletion.

We first examine critical limitations of existing dynamic graph-based ANNS methods~\cite{liu2025wolverine,xu2025place,singh2021freshdiskann}: the latency gap between deletion and insertion, and the loss of query recall as updates accumulate. We use a sliding-window quantification (SWQ) workload~\cite{liu2025wolverine}: the index is built from an initial window containing 70\% of the base vectors; in each subsequent round, 1\% of the vectors expire and are deleted, while the same number of new vectors are inserted, keeping the index size fixed. The workload runs for 50 rounds, and we measure insertion and deletion latencies in each round. We evaluate the in-memory Vamana update method of FreshDiskANN~\cite{singh2021freshdiskann}, denoted FreshVamana (also, IP-Vamana~\cite{xu2025place}). On Deep10M~\cite{babenko2016efficient}, insertions into HNSW~\cite{malkov2018efficient} and Vamana~\cite{jayaram2019diskann} have average latencies of $0.022$ and $0.072$\,ms, whereas deletions with Wolverine~\cite{liu2025wolverine}, FreshVamana, and IP-Vamana~\cite{xu2025place} take $0.048$, $1.009$, and $0.217$\,ms, respectively. Relative to insertion on their underlying graph, the deletion-to-insertion ratios are $2.17\times$, $14.03\times$, and $3.02\times$; the corresponding ratios on MSong~\cite{bertin2011million} are $2.02\times$, $9.41\times$, and $1.86\times$. Figure~\ref{fig:update-latency-motivation} therefore exposes a persistent update asymmetry across different methods.

\begin{table}[t]
  \centering
  \small
  \figureTopMargin
  \caption{Recall@10 under accumulated deletions on Sift1M.}
  \vspace{-3ex}
  \label{tab:recovery-recall}
  \begin{tabular*}{\columnwidth}{@{\extracolsep{\fill}}lccc@{}}
    \toprule
    \textbf{Method}                   & \textbf{Before} & \textbf{30\% deleted} & \textbf{50\% deleted} \\
    \midrule
    Wolverine~\cite{liu2025wolverine} & 0.9962          & 0.9932                & 0.9883                \\
    IP-Vamana~\cite{xu2025place}      & 0.9946          & 0.9967                & 0.9975                \\
    \bottomrule
  \end{tabular*}
  \tableBelowMargin
\end{table}

We further evaluate Recall@10 under cumulative deletions on Sift1M. As shown in Table~\ref{tab:recovery-recall}, Wolverine's recall decreases from $0.9962$ to $0.9883$ after $50\%$ of the vectors are deleted, whereas IP-Vamana maintains stable recall throughout the deletion schedule. The effectiveness of in-place repair depends on whether it identifies the affected in-neighbors and restores their local connections; incomplete recovery leaves search paths unrepaired. \textit{These observations show that existing methods struggle to meet the two requirements simultaneously: 1) making deletion as efficient as insertion, and 2) maintaining stable query recall after sustained updates.}

\section{Analysis on Dynamic Graph-Based ANNS}\label{sec:challenges}

\subsection{Key Challenges}

\begin{figure}[t]
  \centering
  \figureTopMargin
  \includegraphics[width=\linewidth]{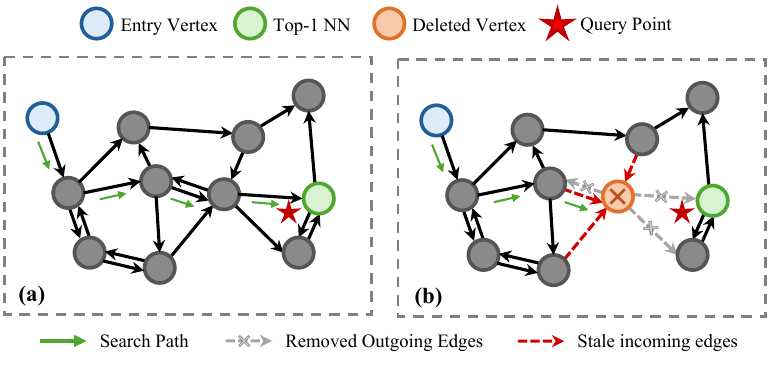}
  \figureCaptionMargin
  \vspace{-6ex}
  \caption{A toy example illustrating the challenges of graph-based ANNS deletion.}
  \label{fig:deletion-challenges}
  \figureBelowMargin\vspace{-2ex}
\end{figure}

The first challenge is identifying the vertices affected by a deletion. An insertion can be completed by searching the graph for candidate neighbors and then adding bidirectional links between the new vertex and these candidates. Deletion, however, must first identify the reverse side of the graph affected by the removed vertex. After deleting $v_d$, the system has to repair vertices that use $v_d$ as an outgoing neighbor, namely the in-neighbor set
\[
  I(v_d)=\{u\in V:\, v_d\in N(u)\}.
\]
Their adjacency lists may still contain stale edges to $v_d$ and therefore need to either remove the obsolete edge or introduce replacement edges to promote local connectivity. As illustrated in Figure~\ref{fig:deletion-challenges}, the outgoing edges of a deleted vertex can be removed immediately, but incoming edges from other vertices may remain stale. Unfortunately, $I(v_d)$ is not stored explicitly by graph-based ANNS indexes. Recovering it exactly requires a full scan over the graph, whereas maintaining it explicitly increases memory consumption and turns each update into a bidirectional adjacency-maintenance operation.

The second challenge is repairing the structural hole left by deletion. Graph search is effective because the index contains both local proximity edges and long-range routing shortcuts. A deleted vertex $v_d$ may serve as a bridge between nearby regions or as a hub traversed by many beam-search trajectories, as shown in Figure~\ref{fig:deletion-challenges}. Once $v_d$ is removed, a search path from the entry point $v_{ep}$ to the nearest neighbor $v_n$ of query $v_q$ may be broken if it relies on the search path $v_{ep}\leadsto v_d\leadsto v_n$, thereby changing the search result. Repeated deletions can gradually fragment the navigable structure even when all remaining vectors are unchanged. Thus, filtering deleted vertices only from final answers is insufficient. The index must promote sufficient local connectivity for subsequent searches to reach the same regions with comparable computational cost.

These two challenges impose conflicting system requirements. Exact reverse-edge discovery and aggressive structural repair improve index accuracy, but they increase deletion latency. Lazy deletion reduces update latency, but it allows the graph topology to degrade over time. \textit{A practical dynamic graph index therefore needs a deletion mechanism that takes effect immediately, avoids explicit incoming-edge maintenance, and confines repair to a sufficiently local region for online serving.}

\subsection{Limitations of Existing Deletion Strategies}

\subsubsection{Why Recall Degrades.}

Lazy-deletion and batch-cleanup methods, including FreshDiskANN~\cite{singh2021freshdiskann}, defer physical deletion. A deleted vertex is marked invalid, but its adjacency records and the reverse edges pointing to it remain in the graph until a later consolidation. Query-time filtering prevents such vertices from being returned as final answers, but it does not remove them from the search process. They may still be visited and inserted into the candidate set. Under a fixed candidate budget $ef_{\mathit{search}}$, each stale candidate consumes capacity that could have been used to explore a live vertex, and each invalid expansion introduces additional distance computations. This effect compounds over long update streams. As stale vertices and stale edges accumulate, the logical graph observed by queries becomes increasingly different from the current live dataset. The search algorithm traverses paths constructed for an older dataset, whereas the result set is evaluated against the current dataset. Periodic consolidation can reset the index state, but the system then faces a tradeoff between quality degradation and expensive rebuilding rather than maintaining stable online behavior.

\begin{figure}[t]
  \centering
  \figureTopMargin
  \begin{subfigure}{0.23\textwidth}
    \centering
    \includegraphics[width=\linewidth]{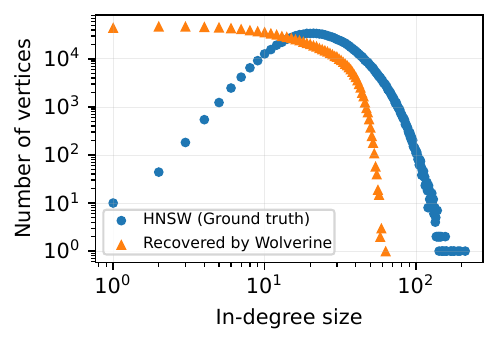}
    \caption{HNSW}
  \end{subfigure}
  \begin{subfigure}{0.23\textwidth}
    \centering
    \includegraphics[width=\linewidth]{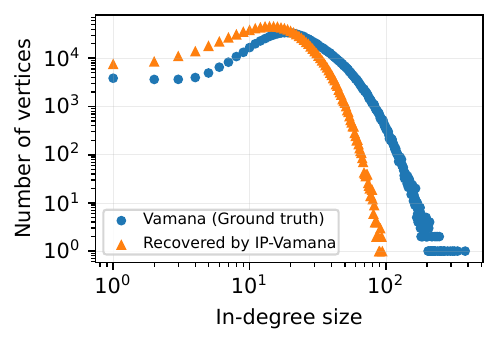}
    \caption{Vamana}
  \end{subfigure}
  \figureCaptionMargin
  \vspace{-1ex}
  \caption{Distribution of original in-degree and recovered in-neighbors under different reverse-edge discovery methods.}
  \label{fig:deletion-impact}
  \figureBelowMargin
\end{figure}

In-place methods approximate the affected in-neighbors. Wolverine~\cite{liu2025wolverine} uses two-hop discovery, while IP-Vamana~\cite{xu2025place} uses graph search. We evaluate their ability to recover the in-neighbor set $I(v_d)$ of a deleted vertex $v_d$ by comparing each vertex's original in-degree with the number of recovered in-neighbors. On Sift1M, the average recovered-to-original in-degree ratios are $0.421$ and $0.664$, respectively (Figure~\ref{fig:deletion-impact}). The higher recovery ratio of IP-Vamana is consistent with its more stable recall in Table~\ref{tab:recovery-recall}. Recovery is particularly poor for high-in-degree vertices due to the hubness effect in high-dimensional spaces~\cite{tomasev2013role}. Missing in-neighbors leave stale incoming edges that consume candidate set capacity and degrade search efficiency and recall, causing recall to continue declining after deletion. \textit{These observations suggest that deletion repair should focus on restoring the local routing topology while providing a mechanism to safely handle unrecovered stale incoming edges. Motivated by this observation, \merit uses version mismatches to invalidate stale incoming edges missed during recovery.}

\begin{figure*}[t]
  \centering
  \figureTopMargin
  \includegraphics[width=0.95\textwidth]{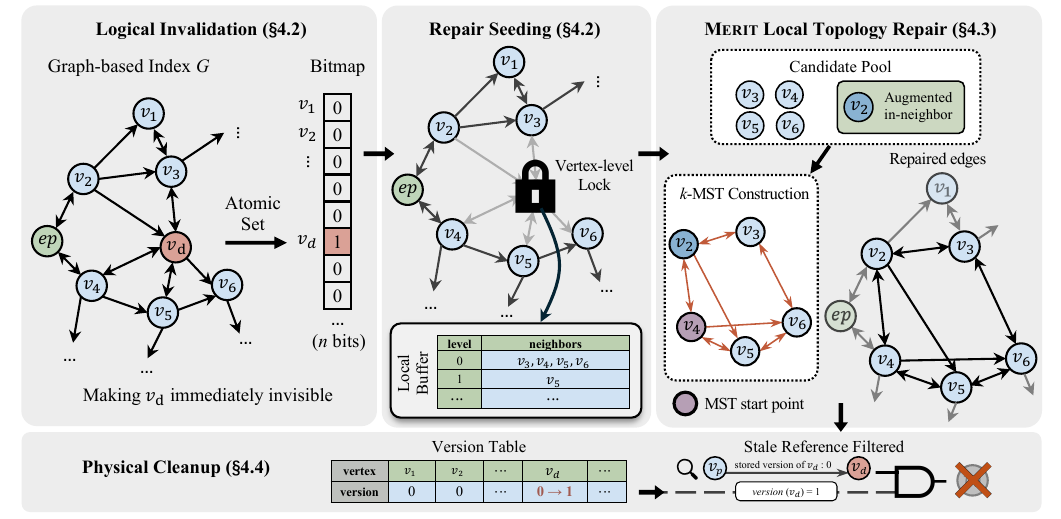}
  \figureCaptionMargin
  \vspace{-1ex}
  \caption{The three algorithmic components of \merit deletion: (1) logical invalidation and local repair seeding for the deleted vertex; (2) approximate construction of the affected repair candidate set and $k$-MST repair over the local candidate set; and (3) versioned-edge invalidation for residual reverse edges.}
  \label{fig:delete-process}
  \figureBelowMargin
\end{figure*}

\subsubsection{Why Deletion Latency Is High.}

In graph-based indexes, insertion is dominated by one graph search followed by local neighbor selection. With maximum out-degree $M$, candidate-list size $L$, and index size $n$, its cost is commonly characterized as $O(M\log n + ML)$~\cite{wang2021comprehensive,fu2017fast}. However, deletion is more involved. The system must identify vertices whose neighbor lists contain the deleted vertex $v_d$, remove these stale incoming edges, determine the affected neighborhood, and repair local connectivity for the involved vertices. Although a deletion can be made almost instantaneous by simply marking $v_d$ as invalid, such lazy deletion leaves the graph structure stale. Existing SOTA methods~\cite{liu2025wolverine,singh2021freshdiskann,xu2025place} therefore perform structural repair during deletion, but this design introduces extra computational overhead.

Naive exact reverse-edge cleanup scans adjacency lists and therefore costs $O(|E|)$ per deletion, which is unacceptable for large online indexes. FreshDiskANN~\cite{singh2021freshdiskann} avoids this cost during individual deletions through lazy updates and deferred batch cleanup. However, during cleanup it examines pairs formed by the outgoing and incoming neighbors of deleted vertices, leading to an average update cost of $O(M^2)$, where $M$ is the maximum out-degree. IP-Vamana~\cite{xu2025place} introduces search-based cleanup to avoid a full graph scan for each deletion, but it pays for additional graph searches and incurs a repair cost of $O(M\log n + c(L+M)(\log L+M))$, where $L$ is the search candidate-list size and $c$ is a hyperparameter. Wolverine~\cite{liu2025wolverine} collects two-hop neighbors of the deleted vertex and heuristically selects valuable repair targets, with repair cost $O(M + M\min(M^2+M, C_s) + M\theta_{\mathrm{in}})$, where $C_s$ and $\theta_{\mathrm{in}}$ are hyperparameters. These repair procedures may still miss affected vertices while imposing substantial online work, making deletion latency significantly higher than insertion latency.

\section{The \merit Algorithm}\label{sec:merit}

With the goal of making deletion as efficient as insertion while maintaining stable recall, we propose \merit, an in-place update mechanism for dynamic graph-based ANNS. \merit is designed to make deleted vertices immediately invisible to search and to repair the affected local topology without explicitly maintaining incoming-neighbor lists.

\subsection{Overview}

The design of \merit follows three principles: (1) locality: each update operation should touch only a small and bounded neighborhood centered at the deleted vertex; (2) immediacy: the deleted vertex should become invisible to subsequent searches as soon as it is marked invalid; and (3) structural stability: long-term deletion and insertion operations should keep graph quality within a stable range rather than cause progressive degradation. Guided by these principles, \merit decomposes each deletion into three ordered components with distinct responsibilities, as illustrated in Figure~\ref{fig:delete-process}.

\begin{enumerate}[leftmargin=1.2em]
  \item \textbf{Logical invalidation and repair seeding.} We mark $v_d$ invalid so that subsequent operations (e.g., searches, insertions and deletions) ignore it, and collect its outgoing neighbors as the initial seed set for local repair (\S\ref{subsec:invalidate}).

  \item \textbf{\merit local topology repair.} We first approximate the missing in-neighbors of $v_d$ by a quick bounded search and merge them with the outgoing-neighbor seed set. We then run an incremental MST-style construction over the local candidate set and attach each candidate with up to $k_r$ short edges under the graph degree bound. The neighborhood snapshot preserves the repair evidence while keeping this stage local and bounded (\S\ref{subsec:repair}).

  \item \textbf{Versioned-edge invalidation.} We remove the residual information of $v_d$ from the graph-based index and increment $\mathrm{version}(v_d)$, ensuring that any remaining stale incoming edges are automatically discarded upon their next access (e.g., during search, insertion, or deletion) due to a version mismatch (\S\ref{subsec:versioning}).
\end{enumerate}


\subsection{Logical Invalidation and Repair Seeding}
\label{subsec:invalidate}

Deleting a vertex from a proximity graph creates two competing requirements. The vertex should immediately cease to participate in subsequent queries and graph updates, whereas its existing neighborhood should remain temporarily available for identifying and repairing the affected region. Physically removing the vertex and its edges at the beginning would destroy useful structural evidence before repair is performed. \merit therefore separates logical invalidation from physical removal.

Given a vertex $v_d$ to be deleted, \merit first marks it as logically invalid. Once this state becomes visible, $v_d$ is excluded from search results, search expansion, repair candidates, and newly constructed adjacency lists. In particular, repair cannot introduce new edges incident to $v_d$, and an adjacency list rewritten during repair filters out invalid vertices. Edges that still reference $v_d$ after invalidation are treated as stale structural records.

This separation gives deletion a well-defined visibility point without requiring all incident edges to be removed immediately. Once the invalidation state is visible, graph operations filter $v_d$ even if an adjacency list still contains its identifier. They also check the vertex state before using $v_d$ as an expansion vertex, returning it as a result, or inserting a new incident edge. Consequently, the logical state of a vertex, rather than the immediate absence of every stale incoming edge, determines its eligibility to participate in the graph.

After invalidating $v_d$, \merit snapshots its outgoing neighborhood, $N(v_d)=\textsc{GetNeighbors}(G,v_d)$, before physically detaching it. The snapshot preserves the local topology that would otherwise be destroyed by other online operations (e.g., insertions and deletions) and provides the direct seeds for subsequent repair. Thus, $v_d$ and its neighborhood remain available as evidence for locating the affected region, while logical invalidation prevents $v_d$ from participating in the repaired graph. Starting from these degree-bounded seeds, \textsc{MeritRepair} discovers affected vertices through local search and incoming-edge inspection, which will be described in \S\ref{subsec:repair}.

\begin{algorithm}[t]
  \caption{\merit-\textsc{Delete}($id$)}
  \label{alg:delete}
  \begin{algorithmic}[1]
    \Require Vertex key $id$, index $\mathcal{I}=\langle V,G,M\rangle$, entry point $ep$, repair parameters $ef'_c,k_r$
    \Ensure Updated graph $G'$
    \State $v_d \gets \textsc{Lookup}(\mathcal{I}, id)$
    \If{$v_d=\bot$ \textbf{or} $\textsc{Invalid}(v_d)$}
    \State \Return $G$
    \EndIf
    \State Mark $v_d$ as logically invalid
    \State $N(v_d) \gets \textsc{GetNeighbors}(G,v_d)$
    \State $G' \gets \textsc{MeritRepair}(G, v_d, N(v_d), ep, ef'_c, k_r)$
    \State Increment $\mathrm{version}(v_d)$
    \State Remove $v_d$ and its outgoing edges from $G'$
    \State \Return $G'$
  \end{algorithmic}
\end{algorithm}

Algorithm~\ref{alg:delete} summarizes the whole deletion process, including logical invalidation, repair seeding, and physical removal. Lines~1--4 resolve the external key and make sure the vertex is valid in the current graph. Line~5 marks $v_d$ as logically invalid and defines the visibility point of deletion. Once this state is observed, $v_d$ is no longer eligible to participate in queries or graph updates. Line~6 snapshots its outgoing neighborhood as the initial repair seed. Line~7 invokes \textsc{MeritRepair}, which expands the affected set through bounded local search and incoming-edge discovery, and then reconstructs local connections efficiently using a $k$-MST repair graph. Finally, Lines~8--9 advance the version of $v_d$ and physically remove its remaining outgoing records. This ordering preserves the pre-deletion topology long enough to guide repair while preventing the deleted vertex from re-entering the active graph.

\subsection{\merit Local Repair}\label{subsec:repair}

The goal of local repair is not to reconstruct every edge incident to $v_d$, but to preserve the routes that used to pass through it. Such a route has the form $u\rightarrow v_d\rightarrow w$, where $u$ is an in-neighbor of $v_d$ and $w$ is an out-neighbor. After deleting $v_d$, the former may lose its next hop, whereas the latter may lose the incoming access that made it reachable from the surrounding graph. Repairing only one side cannot bridge this broken route: using only out-neighbors leaves the vertices that pointed to $v_d$ disconnected from the repair, while using only in-neighbors provides no replacement destinations for their removed edges. We therefore repair the union of the two boundary sets,
\[
  C = \bigl(N(v_d)\cup \widehat{I}(v_d)\bigr)\setminus\{u\mid\textsc{Invalid}(u)\},
\]
where $N(v_d)$ is available from the adjacency snapshot, $\widehat{I}(v_d)$ is an approximate recovery of the in-neighbors, and the function $\textsc{Invalid}(u)$ filters out invalid vertices. With this candidate set, we can construct a local repair graph that preserves connectivity and routing diversity:

\begin{definition}[Local repair graph]
  For a deletion of $v_d$, the local repair graph $H=(C,E^R)$ contains the surviving out-neighbors and recovered in-neighbors in $C$, and the repair edges $E^R$ inserted among them after $v_d$ is removed.
\end{definition}

The graph does not explicitly store $I(v_d)$, and finding it exactly would require a reverse-edge index or a scan of all adjacency lists. Instead, we query the graph with $\vec{x}_{v_d}$ using a small beam width $ef'_c$. Vertices that are easy to reach near $v_d$ are precisely the ones most likely to participate in ordinary search paths through its neighborhood. We retain a returned vertex only if its adjacency list actually contains $v_d$; thus, metric proximity proposes candidates, while the edge test certifies that each retained vertex is an in-neighbor. The recovery is intentionally approximate. Its purpose is to cover the readily searchable part of $I(v_d)$ that affects near-term navigation, not to enumerate all reverse edges, which are safely invalidated by the version mechanism in \S\ref{subsec:versioning}. Because graph-search work grows with its search-list width, lowering $ef'_c$ directly bounds the recovery cost~\cite{peng2023efficient, wang2021comprehensive, wu2026fgim}. This turns reverse-neighbor discovery into one fast, bounded search rather than a global cleanup.
 
Once $C$ is fixed, the repair objective is to promote local connectivity with short edges. This motivates an MST candidate backbone.

\begin{definition}[MST repair]
  For non-empty $C$, let $K(C)$ be the complete weighted graph on $C$ with distance function $\delta(u,v)$, and let $T$ be an MST of $K(C)$. MST repair inserts the bidirectional edges of $T$ into the index, forming a minimum-weight connected backbone over $C$.
\end{definition}

Although an MST guarantees connectivity, connectivity alone is insufficient for ANNS. An MST provides only one path between any pair of candidates and therefore offers few alternative transitions when graph search reaches a local branch. These tree-like bottlenecks can limit beam-search expansion even when the vertices remain reachable~\cite{muja2009flann}. Effective local repair should therefore preserve connectivity while providing multiple short routes for graph search. To achieve this goal, we augment the MST backbone with a few nearby edges, yielding $k_r$-MST repair, which retains the connectivity guarantee while improving routing diversity.

\begin{figure}[t]
  \centering
  \figureTopMargin
  \includegraphics[width=\linewidth]{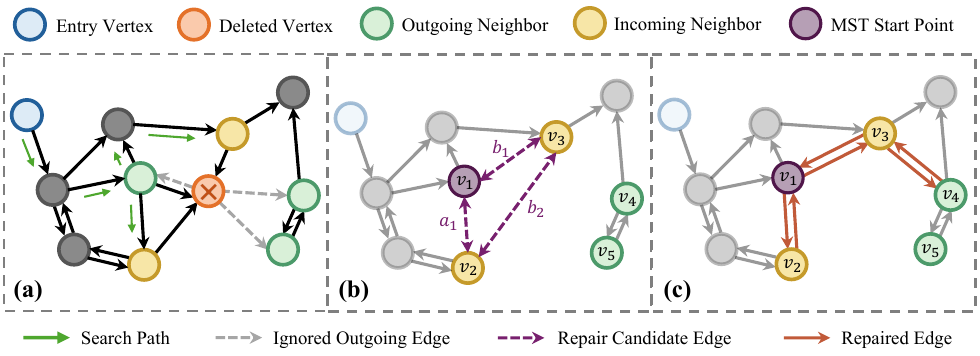}
  \figureCaptionMargin
  \vspace{-4ex}
  \caption{A toy example of \merit local repair. (a) Search-based recovery of repair candidates. (b) Incremental $k_r$-MST construction with $k_r=2$. (c) The repaired local topology.}
  \label{fig:merit-repair}
  \figureBelowMargin
\end{figure}

\begin{table}[t]
\centering
\caption{Comparison of different repair strategies on Gist1M after deleting 50\% of vectors. Search pool size $L$ is fixed at $200$. QPS is measured at Recall@10=$0.93$.}
\vspace{-2ex}
\label{tab:why-kr-mst}
\footnotesize
\setlength{\tabcolsep}{6pt}
\begin{tabular*}{\columnwidth}{@{\extracolsep{\fill}}lcccc@{}}
\toprule
& \multicolumn{2}{c}{Vertex Reachability}
& \multicolumn{2}{c}{Search Quality}\\
\cmidrule(lr){2-3}\cmidrule(lr){4-5}
Strategy
& Before & After
& Recall@10 & QPS\\
\midrule
IP-Vamana~\cite{xu2025place}
&98.65\%&97.70\%
&0.9720&740.02\\

Wolverine~\cite{liu2025wolverine}
&99.22\%&91.68\%
&0.8896&333.21\\

MST
&99.22\%&97.71\%
&0.9656&761.88\\

\textbf{$k_r$-MST (Ours)}
&\textbf{99.22\%}&\textbf{97.90\%}
&\textbf{0.9723}&\textbf{837.91}\\
\bottomrule
\end{tabular*}
\tableBelowMargin
\end{table}

\begin{definition}[$k_r$-MST repair]
  Given the MST backbone $T$ on $C$, a $k_r$-MST repair graph $H$ is obtained by attaching each candidate to up to $k_r$ closest neighbors in the already connected candidate set, including its MST parent, subject to the graph degree bound $M$.
\end{definition}

For any non-empty $C$ and $k_r\ge1$, $k_r$-MST preserves the connectivity of the local repair graph after RNG-based pruning. The repair graph is constructed with the MST backbone $T$ and adds up to $k_r-1$ additional nearby edges for each vertex, providing bounded routing redundancy. Since any MST of $C$ is contained in the RNG of $C$~\cite{toussaint1980relative}, i.e., $T\subseteq\operatorname{RNG}(C)$, RNG pruning removes only redundant repair edges while preserving the MST backbone. As the MST edges are inserted as bidirectional connections, the preserved backbone maintains reachability of the repaired candidate set. Therefore, \textsc{InsertAndPrune} (see Algorithm~\ref{alg:repair}) retains connectivity while the additional edges improve branching and path diversity during search.

We further evaluate this design by comparing MST repair with $k_r$-MST and representative repair methods, including IP-Vamana~\cite{xu2025place} and Wolverine~\cite{liu2025wolverine}, on Gist1M after deleting $50\%$ of vectors. As shown in Table~\ref{tab:why-kr-mst}, $k_r$-MST achieves the highest vertex reachability and search quality among the compared repair strategies. In contrast, MST preserves reachability but provides inferior search performance. These results validate our design intuition that connectivity preservation alone is insufficient for ANNS. The MST backbone provides reachability, while the additional bounded edges in $k_r$-MST introduce routing alternatives that improve search quality.

We choose the closest candidate to the search entry point $ep$ as the $k_r$-MST start point. This choice biases the repaired structure toward the direction from which graph search is likely to enter the local region, allowing the search to expand progressively into the tree while avoiding an unnecessarily indirect initial access to the repaired candidates. We use this as a navigation heuristic rather than claiming that the resulting tree preserves the original routes or minimizes search-path length.

Figure~\ref{fig:merit-repair} illustrates the complete process. In Figure~\ref{fig:merit-repair}(a), search paths from the entry vertex reach vertices around $v_d$; the adjacency test distinguishes true recovered in-neighbors from merely nearby vertices, and the ignored outgoing edges of $v_d$ are already available from its snapshot. Figure~\ref{fig:merit-repair}(b) shows the incremental construction for $k_r=2$. The construction starts from $v_1$, the candidate closest to the entry point. The first candidate $v_2$ can create only $a_1=(v_2,v_1)$ because the current tree contains one vertex. After $v_2$ joins the tree, $v_3$ can create two candidate edges, $b_1=(v_3,v_1)$ and $b_2=(v_3,v_2)$. Figure~\ref{fig:merit-repair}(c) shows the repaired local topology after the same rule is applied to the remaining candidates.

\begin{algorithm}[t]
  \caption{\merit: local candidate construction and repair}
  \label{alg:repair}
  \begin{algorithmic}[1]
    \Require Deleted vertex $v_d$ with vector $\vec{x}_{v_d}$, neighborhood snapshot $N(v_d)$, graph $G$, entry $ep$, beam width $ef'_c$, repair degree $k_r$, degree bound $M$
    \Ensure Updated local topology of $G$
    \State $\mathcal{R} \gets \textsc{KnnSearch}(\vec{x}_{v_d}, G, ef'_c, ef'_c, ep)$
    \State $\mathcal{I} \gets \{u\in \mathcal{R} \mid \neg\textsc{Invalid}(u)\wedge v_d\in N(u)\}$
    \State $C \gets \{u\in N(v_d)\mid\neg\textsc{Invalid}(u)\}\cup \mathcal{I}$
    \If{$C=\emptyset$}
    \State \Return
    \EndIf
    \State $s \gets \arg\min_{u\in C}\delta(u,ep)$
    \State $T \gets \{s\}$, $U \gets C\setminus\{s\}$
    \While{$U\neq\emptyset$}
    \State $(u,w^\star) \gets \arg\min_{(u,w)\in U\times T}\delta(u,w)$
    \State $P \gets$ $k_r$ closest vertices to $u$ in $T$
    \For{$w\in P$}
    \State \textsc{InsertAndPrune}$(G,u,w,M)$
    \State \textsc{InsertAndPrune}$(G,w,u,M)$
    \EndFor
    \State $T \gets T\cup\{u\}$, $U\gets U\setminus\{u\}$
    \EndWhile
    \State \Return
  \end{algorithmic}
\end{algorithm}

Algorithm~\ref{alg:repair} implements the two stages for the single-layer graph used in our description. Lines~1--3 construct the repair candidates. Line~1 runs the bounded search with $\vec{x}_{v_d}$ and returns the pool $\mathcal{R}$. Line~2 filters this pool to live vertices whose adjacency lists contain $v_d$, producing $\mathcal{I}=\widehat{I}(v_d)$. Line~3 unions these recovered in-neighbors with the outgoing-neighbor snapshot $N(v_d)$. Lines~4--17 construct the $k_r$-MST repair graph. Lines~4--6 make an empty candidate set a no-op. Line~7 chooses the candidate closest to $ep$ as the construction start, and Line~8 partitions the candidates into the connected set $T$ and unconnected set $U$. Lines~9--17 repeat until every candidate is attached. Line~10 applies the Prim rule, selecting the minimum-distance pair across the cut $(U,T)$. Line~11 chooses up to $k_r$ closest vertices to $u$ in $T$; when $|T|<k_r$, it simply uses all available tree vertices, as for edge $a_1$ in Figure~\ref{fig:merit-repair}(b). Lines~12--15 insert both directions of every selected edge and prune the affected lists to degree $M$. Line~16 then moves $u$ into $T$.

\subsection{Versioned-Edge Invalidation}\label{subsec:versioning}

Local repair adds candidate routes within the searchable part of the affected neighborhood, but it does not recover every in-neighbor of $v_d$. Consequently, an undiscovered vertex $u$ may still store a stale incoming edge $u\rightarrow v_d$. Finding and removing all such edges would require either explicit in-neighbor lists or periodic scans of the entire graph. \merit avoids both requirements by separating \emph{logical invalidation} from \emph{physical removal}. A version change immediately makes every stale edge to $v_d$ unusable, while ordinary graph operations gradually remove obsolete edge entries from their containing adjacency lists.

To support this separation, each stored edge carries the version of its target vertex at the time the edge is created. The index additionally maintains a global version table $\mathcal{V}$ with one current version $\mathcal{V}[v]$ for every vertex slot $v$. For example, when an adjacency-list element is represented by a 64-bit word, \merit uses the high 16 bits for the target version and the low 48 bits for the target identifier. The resulting encoding is
\[
  e_{u\rightarrow v}
  =
  \underbrace{\mathcal{V}[v]}_{\text{high 16 bits}}\ll 48
  \;|\;
  \underbrace{\operatorname{id}(v)}_{\text{low 48 bits}}.\vspace{-1.5ex}
\]
The version belongs to the target $v$, rather than the source $u$, because one update to $\mathcal{V}[v]$ must invalidate incoming edges to $v$ stored across many different adjacency lists. The 48-bit identifier field supports up to $2^{48}$ vertex slots; other word layouts can use the same scheme with a different bit allocation.

An edge word with decoded fields $(\nu_e,v)$, where $\nu_e$ is the stored target version, is valid only if $\nu_e=\mathcal{V}[v]\wedge\neg\textsc{Invalid}(v)$. Deleting $v_d$ increments $\mathcal{V}[v_d]$ after marking the vertex invalid. Every previously stored edge $u\rightarrow v_d$ contains the preceding version and therefore fails this check, regardless of where the edge resides or whether $u$ was found during local repair. A single change to the version table thus logically invalidates all residual stale incoming edges without locating them individually. This property also prevents a stale edge from becoming valid merely because the physical slot of $v_d$ is later reused, since a new edge to that slot is encoded with its new current version.

During beam search, \merit decodes each edge word and excludes a version-mismatched or invalid target before the target enters the candidate queue or is expanded (i.e., checked in Line~6 of Algorithm~\ref{alg:anns}). Hence, physically retained stale incoming edges consume neither result capacity nor search-expansion budget. The same filtered view is used by graph updates. When insertion, deletion repair, or neighbor pruning rewrites an adjacency list, it copies only currently valid neighbors into the new list. Stale edge entries are omitted, and newly inserted edges are stamped with the targets' current versions. The rewritten list replaces the previous adjacency data, thereby removing any obsolete edge entries it contained. Adjacency lists that are never updated may retain stale edge entries physically, but, before version wrap-around, those entries remain invisible to graph operations. Therefore, continued operation progressively compacts touched lists and requires no periodic full-graph cleanup.

\section{Complexity and Lifetime}\label{sec:analysis}

In this section, we analyze the complexity of \merit and provide a probabilistic bound on the lifetime of the version counter.

\subsection{Complexity Analysis}

In this part, we analyze the expected cost of a single deletion in \merit. The deletion process consists of three main components: invalidation, local repair, and versioned-edge invalidation.

Let $n=|V|$, let $M$ be the maximum out-degree, and let $C$ be the local repair candidate set. Logical invalidation, neighborhood snapshotting, and physical removal only take $O(|N(v_d)|)=O(M)$ time. Recovering the in-neighbor candidates through bounded beam search requires $O(ef'_c\log n)$ distance computations~\cite{wu2026fgim,wang2021comprehensive,peng2023efficient}. The heap-optimized Prim construction~\cite{johnson1975priority} takes $O(|E_C|\log |C|)=O(|C|^2\log |C|)$ time, where $E_C$ is the edge set of the implicit complete graph over $C$, while inserting and pruning up to $k_r$ edges for every candidate takes $O(|C|k_rM)$ time. Notably, incrementing the version counter only takes $O(1)$ time. The deletion cost is therefore
\[
  O\bigl(M+ef'_c\log n+|C|^2\log |C|+|C|k_rM\bigr).
\]
Because $|C|\le M+ef'_c$, the repair cost depends only on configured local bounds rather than unbounded reverse degree. With $ef'_c=cM$ for a small constant $c$ and fixed $k_r$, the total cost is $O(M\log n+M^2\log M)$. Unlike prior update methods~\cite{liu2025wolverine, xu2025place}, this analysis also includes the cost of handling residual stale incoming edges, which constitute a significant component of deletion cost. For the space complexity, the graph occupies $O(nM)$ space, while the version table and deletion bitmap add $O(n)$ space.

\subsection{Lifetime of the Version Counter}\label{subsec:lifetime}

The worst case is deterministic: a physical slot wraps after $V_{\max}=2^{16}=65\,536$ reuses, after which a sufficiently old edge could exhibit the ABA problem~\cite{dechev2010understanding}. For context, we also give a probability bound under an \emph{idealized uniform-reuse model}: each of $U$ increments independently selects one of $n$ slots uniformly. Then $X_i\sim\mathrm{Binomial}(U,1/n)$ with mean $\mu=U/n$. For $\mu<V_{\max}$, a union bound and Chernoff bound~\cite{motwani1996randomized} give
\begin{equation}
  \begin{aligned}
     & g(\mu)=V_{\max}\ln\!\frac{V_{\max}}{\mu}-V_{\max}+\mu, \\
     & \Pr\!\left[\max_i X_i\ge V_{\max}\right]
    \le n\exp[-g(\mu)].
  \end{aligned}
  \label{eq:wrap-bound}
\end{equation}
Solving $n\exp[-g(\mu)]\le\delta$ gives a model-dependent safe budget $U=n\mu$. For example, $n=10^8$ and $\delta=10^{-6}$ give $\mu_\delta\approx63\,502$ and $U_{\mathrm{safe}}\approx6.35\times10^{12}$ version increments, corresponding to about 20.1 years at $10^4$ increments per second. This budget suggests that the 16-bit version encoding is sufficiently durable for long-running industrial deployments under the assumed highly dynamic workload.

\begin{table}[t]
  \centering
  \small
  \caption{Evaluation Datasets.}
  \vspace{-3ex}
  \label{tab:dataset}
  \begin{tabular}{lcccc}
    \toprule
    \textbf{Dataset}                     & \textbf{Dim} & \textbf{Size}   & \textbf{\#Queries} & \textbf{Metric} \\
    \midrule
    Sift1M~\cite{jegou2010product}       & 128          & 1{,}000{,}000   & 10{,}000           & $\ell_2$        \\
    Gist1M~\cite{jegou2010product}       & 960          & 1{,}000{,}000   & 1{,}000            & $\ell_2$        \\
    Deep1M~\cite{babenko2016efficient}   & 96           & 1{,}000{,}000   & 10{,}000           & cosine          \\
    Deep10M~\cite{babenko2016efficient}  & 96           & 10{,}000{,}000  & 10{,}000           & cosine          \\
    Deep100M~\cite{babenko2016efficient} & 96           & 100{,}000{,}000 & 10{,}000           & cosine          \\
    GloVe~\cite{pennington2014glove}     & 100          & 1{,}183{,}513   & 10{,}000           & cosine          \\
    MSong~\cite{bertin2011million}       & 420          & 994{,}185       & 1{,}000            & $\ell_2$        \\
    \bottomrule
  \end{tabular}
  \tableBelowMargin
\end{table}

\begin{figure*}[t!]
  \centering
  \includegraphics[width=.46\textwidth]{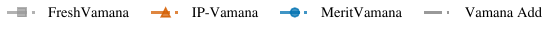}\\[-.35em]
  \resultsixpanel{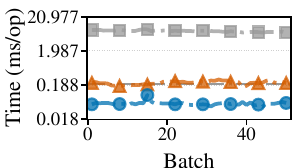}{Sift1M, 5\%.}\hfill
  \resultsixpanel{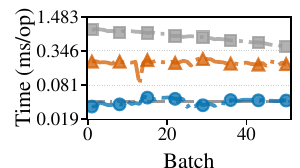}{Sift1M, 50\%.}\hfill
  \resultsixpanel{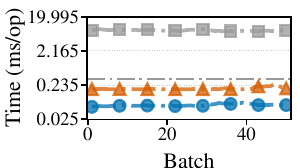}{Deep1M, 5\%.}\hfill
  \resultsixpanel{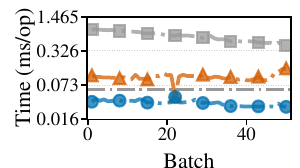}{Deep1M, 50\%.}\hfill
  \resultsixpanel{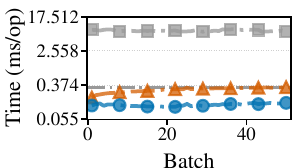}{GloVe, 5\%.}\hfill
  \resultsixpanel{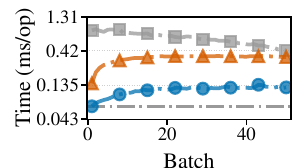}{GloVe, 50\%.}
  \figureCaptionMargin
  \vspace{-1ex}
  \caption{Amortized deletion time for Vamana-based methods. The gray dashed
    line marks the Vamana insertion time.}
  \label{fig:rq1-vamana}
\end{figure*}

\begin{figure*}[t!]
  \centering
  \vspace{-3ex}
  \includegraphics[width=.24\textwidth]{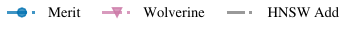}\\[-.4em]
  \resultsixpanel{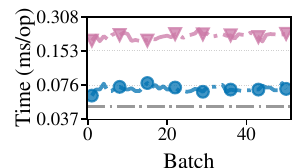}{Sift1M, 5\%.}\hfill
  \resultsixpanel{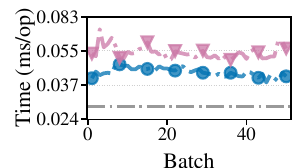}{Sift1M, 50\%.}\hfill
  \resultsixpanel{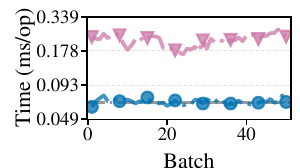}{Deep1M, 5\%.}\hfill
  \resultsixpanel{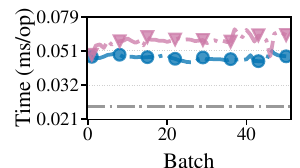}{Deep1M, 50\%.}\hfill
  \resultsixpanel{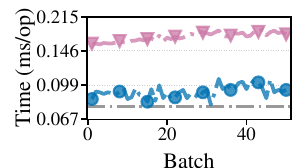}{GloVe, 5\%.}\hfill
  \resultsixpanel{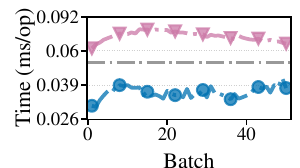}{GloVe, 50\%.}
  \figureCaptionMargin
  \vspace{-1ex}
  \caption{Amortized deletion time for HNSW-based methods. The gray dashed
    line marks the HNSW insertion time.}
  \label{fig:rq1-graph}
  \figureBelowMargin
\end{figure*}

\begin{figure*}[t!]
  \centering
  \includegraphics[width=.52\textwidth]{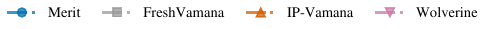}\\[-.35em]
  \resultsixpanel{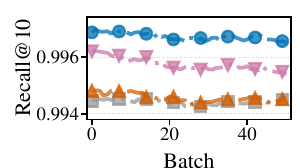}{Sift1M, 5\%.}\hfill
  \resultsixpanel{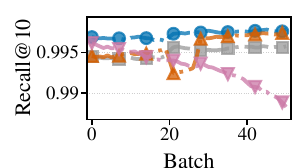}{Sift1M, 50\%.}\hfill
  \resultsixpanel{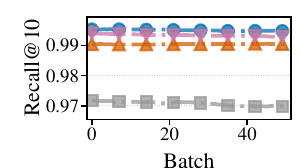}{Deep1M, 5\%.}\hfill
  \resultsixpanel{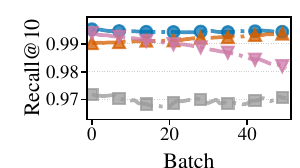}{Deep1M, 50\%.}\hfill
  \resultsixpanel{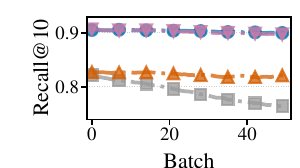}{GloVe, 5\%.}\hfill
  \resultsixpanel{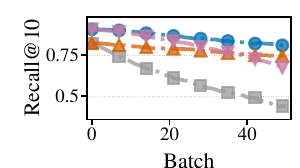}{GloVe, 50\%.}
  \figureCaptionMargin
  \vspace{-1ex}
  \caption{Post-deletion Recall@10 at deletion rates of 5\% and 50\%.}
  \label{fig:rq2-quality}
  \figureBelowMargin
\end{figure*}

\section{Experimental Evaluation}\label{sec:experiments}

Our evaluation answers five research questions.
\begin{enumerate}[topsep=0pt, itemsep=0pt, parsep=0pt]
  \item[\textbf{RQ1}] How efficiently does \merit process deletions on SOTA graph indexes?
  \item[\textbf{RQ2}] How well does \merit preserve search quality as deletions accumulate?
  \item[\textbf{RQ3}] Does \merit jointly provide low update latency and stable search quality under sustained delete--insert replacements?
  \item[\textbf{RQ4}] How do the repair degree $k_r$, versioned-edge invalidation, and MST repair contribute to performance?
  \item[\textbf{RQ5}] How does \merit scale to a 100-million-point dataset?
\end{enumerate}

\subsection{Setup}\label{subsec:setup}

\noindent\textbf{Datasets.} The experiments are conducted on several popular benchmarking datasets. All of them are real-world datasets and have been widely used in the literature~\cite{wang2021comprehensive,aumuller2020ann}. As summarized in Table~\ref{tab:dataset}, the datasets cover images (Sift1M~\cite{jegou2010product}, Gist1M~\cite{jegou2010product}, Deep1M, Deep10M, and Deep100M~\cite{babenko2016efficient}), text (GloVe~\cite{pennington2014glove}), and music (MSong~\cite{bertin2011million}).

\begin{figure*}[t!]
  \centering
  \includegraphics[width=.52\textwidth]{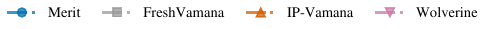}\\[-.35em]
  \resultsixpanela{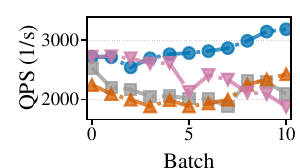}{Sift1M QPS.}\hfill
  \resultsixpanela{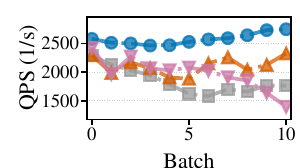}{Deep1M QPS.}\hfill
  \resultsixpanela{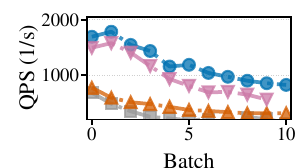}{GloVe QPS.}\hfill
  \resultsixpanela{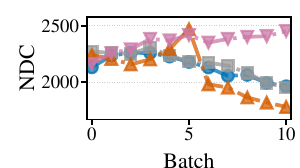}{Sift1M NDC.}\hfill
  \resultsixpanela{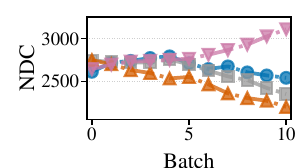}{Deep1M NDC.}\hfill
  \resultsixpanela{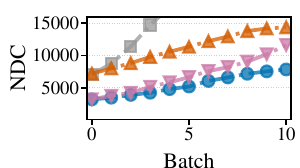}{GloVe NDC.}
  \figureCaptionMargin
  \vspace{-1ex}
  \caption{Search throughput and average distance computations per query
    after each deletion batch at $r=50\%$.}
  \label{fig:rq2-efficiency}
  \figureBelowMargin
\end{figure*}

\begin{figure*}[t!]
	\centering
	\figureTopMargin
	\includegraphics[width=.52\textwidth]{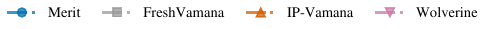}\\[-.35em]
	\resultsixpanela{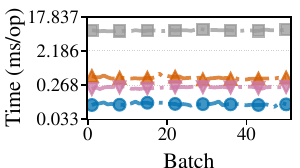}{Sift1M, 5\%.}\hfill
	\resultsixpanela{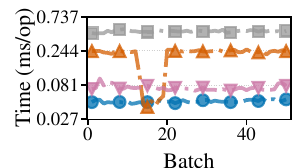}{Sift1M, 50\%.}\hfill
	\resultsixpanela{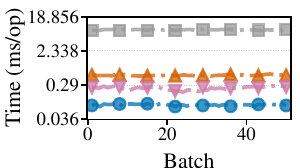}{Deep1M, 5\%.}\hfill
	\resultsixpanela{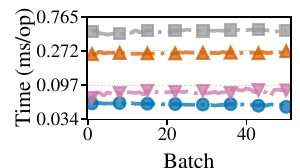}{Deep1M, 50\%.}\hfill
	\resultsixpanela{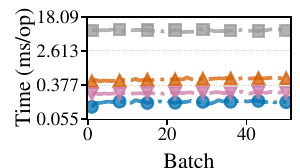}{GloVe, 5\%.}\hfill
	\resultsixpanela{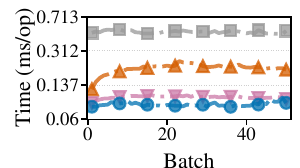}{GloVe, 50\%.}
	\figureCaptionMargin
	\vspace{-1ex}
	\caption{Amortized update time $L_{\rm upd}$ under the
		fixed-cardinality sliding-window update workload.}
	\label{fig:rq3-latency}
	 \figureBelowMargin
\end{figure*}

\begin{figure*}[t!]
	\centering
	\includegraphics[width=.52\textwidth]{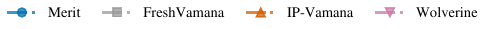}\\[-.35em]
	\resultsixpanela{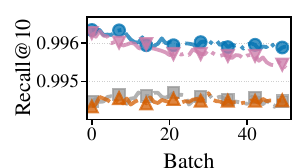}{Sift1M, 5\%.}\hfill
	\resultsixpanela{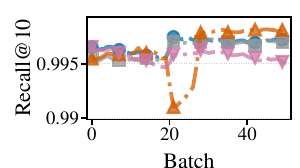}{Sift1M, 50\%.}\hfill
	\resultsixpanela{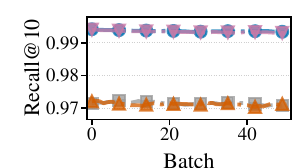}{Deep1M, 5\%.}\hfill
	\resultsixpanela{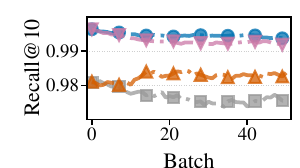}{Deep1M, 50\%.}\hfill
	\resultsixpanela{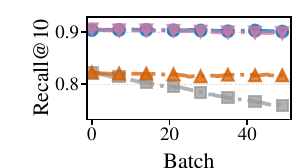}{GloVe, 5\%.}\hfill
	\resultsixpanela{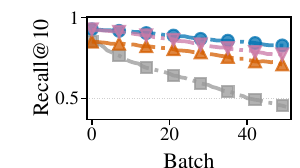}{GloVe, 50\%.}
	\figureCaptionMargin
	\vspace{-1ex}
	\caption{Post-update Recall@10 under the fixed-cardinality sliding-window
		update workload.}
	\label{fig:rq3-recall}
	\figureBelowMargin
\end{figure*}

\noindent\textbf{Algorithms.} The evaluation covers two SOTA graph structures. For HNSW indexes, we compare Wolverine~\cite{liu2025wolverine} with \merit implemented on HNSW. For single-layer Vamana indexes, we compare FreshVamana, the in-memory update algorithm of FreshDiskANN~\cite{singh2021freshdiskann}, and IP-Vamana~\cite{xu2025place}, the in-memory update algorithm of IP-DiskANN, with \merit implemented on Vamana (i.e., MeritVamana).

\noindent\textbf{Parameters.} For a fair comparison, we use the same maximum out-degree, $M=R\in\{32,64\}$, and a search beam width of 200 for all methods. Method-specific parameters follow the settings recommended in the original papers. Unless otherwise stated, \merit uses $k_r=2$, $16$ version bits, and $ef'_c=2M$. Wolverine uses $\theta_{\mathrm{in}}=32$ and $C_s=64$. FreshVamana uses $L=200$ and $\alpha=1.2$, and IP-Vamana additionally uses $l_d=128$, $k=50$, and $c=3$. We do not apply any additional quantization (e.g., product quantization (PQ)~\cite{jegou2010product} or scalar quantization (SQ)~\cite{aguerrebere2023similarity}) to any of the methods, because our focus is on the graph update.

\noindent\textbf{Metrics.} We report Recall@10, QPS, and NDC to evaluate search quality, following the definitions in \S\ref{subsubsec:anns}. We measure update cost as the amortized wall-clock time per operation, i.e., the end-to-end elapsed time of processing a batch divided by the number of operations in that batch. We report deletion latency ($L_{\rm del}$) for deletion workloads. Under the update workload, the amortized update cost is defined as $L_{\rm upd}=L_{\rm ins}+L_{\rm del}$ following \S\ref{subsec:update-model}.

\noindent\textbf{Computing Environment.}
We implemented our method in C++11 and compiled the code using CMake 3.22.1 with GCC 11.4.0 as the compiler. All experiments run on a server with an AMD EPYC 9554 (64 cores / 128 threads) and 377~GB RAM under Ubuntu 22.04. Build and update use 128 threads; search is single-threaded to isolate per-query cost.

\noindent\textbf{Protocol.} The deletion workload removes vertices in $50$ batches at total deletion rates $r\in\{5\%,50\%\}$ and evaluates search after every batch. The sustained-update experiments use the SWQ workload defined in \S\ref{subsec:update-motivation}. Every reported configuration uses identical update sequences across competing methods. Each experiment is repeated three times with the mean result reported.

\subsection{RQ1 -- Deletion Efficiency}\label{subsec:delete-lat}

We begin by measuring amortized deletion time per operation over successive deletion batches. Methods are compared only within the same graph structure, with separate results for Vamana (i.e., MeritVamana) and HNSW implementations.

\noindent\textbf{Answer to RQ1.} As shown in Figures~\ref{fig:rq1-vamana} and~\ref{fig:rq1-graph}, \merit achieves consistently low deletion latency under both removal ratios ($r=5\%$ and $r=50\%$), substantially outperforming existing SOTA methods on three datasets. Specifically, when $r=50\%$, \merit achieves average deletion latencies of $0.0399$, $0.0333$, and $0.1031~\mathrm{ms}$ on Sift1M, Deep1M, and GloVe, respectively. By comparison, FreshVamana requires $0.6459$, $0.6285$, and $0.6529~\mathrm{ms}$, while IP-Vamana requires $0.2110$, $0.1039$, and $0.3363~\mathrm{ms}$ on the same datasets (Figure~\ref{fig:rq1-vamana}). Figure~\ref{fig:rq1-graph} further shows that, under the $50\%$ removal ratio, \merit consistently outperforms Wolverine, reducing the deletion latency from $0.0545$, $0.0585$, and $0.1085~\mathrm{ms}$ to $0.0398$, $0.0396$, and $0.0359~\mathrm{ms}$, respectively. Overall, \merit delivers up to $3.02\times$--$18.87\times$ speedup over existing SOTA approaches. This improvement stems from its ability to rapidly identify candidate neighbors and efficiently repair the affected graph structure, without performing an expensive global cleanup of incoming edges.

\subsection{RQ2 -- Search Quality after Deletions}\label{subsec:query-perf}

In order to evaluate the impact of deletions on search quality, we next examine Recall@10 after every deletion batch at total deletion rates of 5\% and 50\%, using the same removal order and per-rate schedules. Search pool size $L$ is set to 200 for all methods.

\noindent\textbf{Answer to RQ2.} As shown in Figure~\ref{fig:rq2-quality}, \merit maintains consistently high recall under both removal ratios across datasets with varying levels of difficulty, from Sift1M to GloVe. When $r=50\%$, the minimum Recall@10 values achieved by \merit are $0.9956$, $0.9921$, and $0.8473$ on Sift1M, Deep1M, and GloVe, respectively. Furthermore, Figure~\ref{fig:rq2-quality} reports the QPS and NDC of different methods at fixed recall levels under $r=50\%$ (Recall@10=99\% for Sift1M and Deep1M, and Recall@10=80\% for GloVe). \merit exhibits a stable trend throughout the 50 deletion batches. This stability is attributed to \merit’s repair strategy, which constructs a $k_r$-MST over locally affected structures along search paths, restoring graph connectivity while preserving navigability. Although all baseline methods remain relatively stable under a small deletion setting of $r=5\%$, their search quality exhibits noticeable fluctuations or degradation when half of the indexed vectors are removed. Overall, \merit preserves stable recall even under high-frequency deletions.

\subsection{RQ3 -- Sustained Update Performance}\label{subsec:longterm}

We then turn to the fixed-cardinality sliding-window update workload, measuring amortized replacement time together with post-update Recall@10. Reporting both metrics reveals whether a method lowers update cost by sacrificing long-term search quality.

\noindent\textbf{Answer to RQ3.} Figure~\ref{fig:rq3-latency} reports the latency of updating $5\%$ and $50\%$ of the vectors on the three datasets. When $r=50\%$, the HNSW-based implementation of \merit achieves average per-update latencies of $0.0476$, $0.0541$, and $0.1022~\mathrm{ms}$ on three datasets. As shown in Figure~\ref{fig:rq3-recall}, \merit also maintains consistently high and stable Recall@10 after each update batch across all datasets. In contrast, IP-Vamana exhibits substantial recall fluctuations on Sift1M, potentially due to the dataset's non-uniform local density \cite{jegou2010product}. Overall, under workloads involving frequent and large-scale deletions and insertions, \merit achieves lower update latency and more stable recall than existing SOTA methods.

\begin{figure}[t]
  \centering
  \figureTopMargin
  \includegraphics[width=\columnwidth]{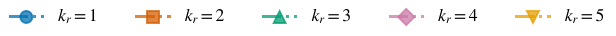}\\[-.3em]
  \resultsmallpanela{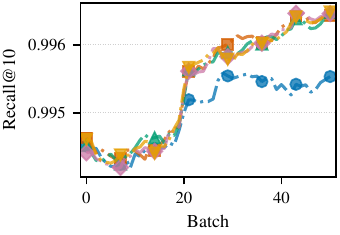}{Sift1M Recall@10.}\hfill
  \resultsmallpanela{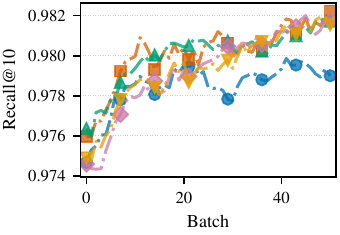}{Gist1M Recall@10.}
  \par\vspace{.5em}
  \hspace{0.5ex}\resultsmallpanela{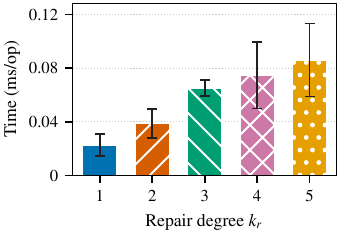}{Sift1M deletion time.}\hfill
  \hspace{1ex}\resultsmallpanela{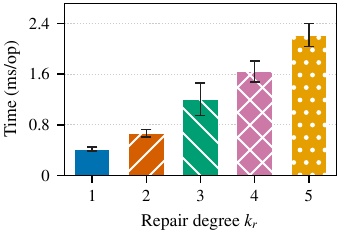}{Gist1M deletion time.}
  \figureCaptionMargin
  \caption{\merit $k_r$ sensitivity.}
  \label{fig:rq4-kr}
\end{figure}

\subsection{RQ4 -- Ablation Study}\label{subsec:ablation}

We study two questions in this ablation. \textbf{RQ4-1} examines how $k_r$ affects amortized deletion time and search quality. \textbf{RQ4-2} separates the effects of MST repair and versioned edges through four component variants.

\noindent\textbf{RQ4-1.} We vary $k_r$ from $1$ to $5$ to evaluate its impact on search quality and average deletion latency. As shown in Figure~\ref{fig:rq4-kr}, increasing $k_r$ incurs higher repair costs, while the corresponding quality gains gradually diminish. On Gist1M, increasing $k_r$ from $1$ to $2$ raises the average deletion latency from $0.41$ to $0.67~\mathrm{ms}$ and improves the final Recall@10 from $0.979$ to $0.982$. When $k_r=5$, the average deletion latency further increases to $2.22~\mathrm{ms}$, while Recall@10 shows no additional improvement. Overall, setting $k_r$ to $2$ or $3$ provides a favorable trade-off by preserving local graph connectivity for each affected node while maintaining low deletion latency.

\begin{figure}[t]
	\centering
	\figureTopMargin\vspace{-2ex}
	\includegraphics[width=.58\columnwidth]{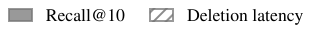}\\[-.2em]
	\begin{subfigure}[t]{.495\columnwidth}
		\centering
		\includegraphics[width=\linewidth]{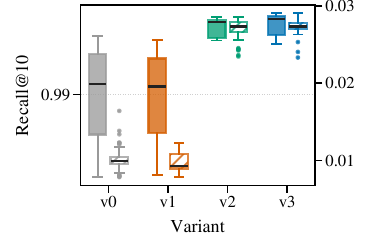}
		\captionsetup{skip=.15em,justification=centering}
		\caption{Sift1M.}
	\end{subfigure}
	\begin{subfigure}[t]{.495\columnwidth}
		\centering
		\includegraphics[width=\linewidth]{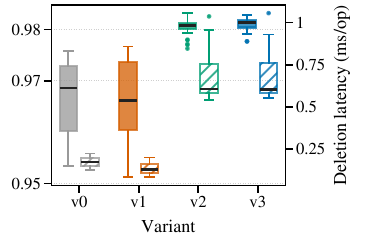}
		\captionsetup{skip=.15em,justification=centering}
		\caption{Gist1M.}
	\end{subfigure}
	\figureCaptionMargin
	\caption{\merit component ablation.}
	\label{fig:rq4-component}
	\figureBelowMargin
	\vspace{-1ex}
\end{figure}

\noindent\textbf{RQ4-2.} We isolate versioned edges and MST repair with four variants. \merit-v0 disables both and performs no repair; \merit-v1 enables only versioned edges and likewise performs no repair; \merit-v2 enables only MST repair; and \merit-v3 enables both. As shown in Figure~\ref{fig:rq4-component}, enabling MST-based repair substantially improves recall while increasing deletion latency (e.g., on Gist1M, it prevents an approximately $3\%$ drop in Recall@10 but incurs a $0.4~\mathrm{ms}$ increase in deletion time). In contrast, versioned edges introduce negligible overhead and avoid the latency spikes incurred by prior methods that periodically remove stale edges from the graph.

\subsection{RQ5 -- Scalability}\label{subsec:scale}

Finally, we examine amortized deletion time and Recall@10 beyond the million-scale setting. The Deep100M run uses the same 50-batch deletion schedule and tuned initial recall.

\begin{figure}[t]
  \centering
  \figureTopMargin
  \includegraphics[width=.5\columnwidth]{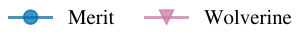}\\[-.35em]
  \begin{subfigure}[t]{.48\columnwidth}
    \centering
    \includegraphics[width=\linewidth]{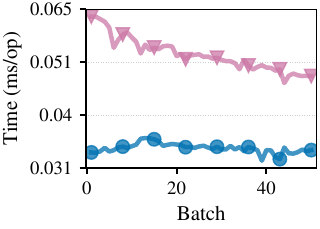}
    \captionsetup{skip=1ex,justification=centering,margin={0pt,20pt}}
    \caption{Deletion time.}
  \end{subfigure}\hfill
  \begin{subfigure}[t]{.48\columnwidth}
    \centering
    \includegraphics[width=\linewidth]{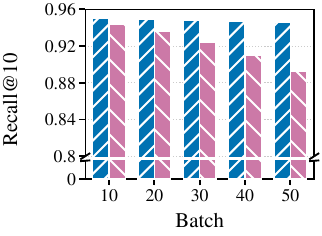}
    \captionsetup{skip=1ex,justification=centering,margin={0pt,20pt}}
    \caption{Recall@10.}
  \end{subfigure}
    \figureCaptionMargin
  \caption{Deep100M scalability over 50 deletion batches.}
    \figureBelowMargin
  \label{fig:rq5}
\end{figure}

\noindent\textbf{Answer to RQ5.} As shown in Figure~\ref{fig:rq5}, \merit achieves an average per-deletion latency of $0.0342~\mathrm{ms}$ on Deep100M and maintains stable latency throughout the experiment. We further compare \merit with Wolverine under the same large-scale deletion workload. Although the two methods start with comparable Recall@10, their search quality diverges as deletions accumulate. After $50$ deletion batches (i.e., removing $50$ million vectors), the Recall@10 of \merit decreases by less than $\sim 1\%$. In contrast, Wolverine exhibits noticeable quality degradation despite repairing the graph after each deletion batch. These results demonstrate that \merit scales effectively to large-scale datasets while preserving both low deletion latency and robust search quality, making it suitable for large-scale production workloads.

\section{Related Work}\label{sec:related}

Graph-based ANNS indexes evolved from k-NN graph construction (e.g., NN-Descent~\cite{dong2011efficient}) and navigation techniques such as NSG and NSWG~\cite{fu2017fast,malkov2014approximate}. Representative in-memory indexes include HNSW~\cite{malkov2018efficient}, which uses a probabilistic hierarchy for skip-list-like navigation~\cite{pugh1990skip}, and Vamana~\cite{jayaram2019diskann}, whose $\alpha$-robust pruning is reported to achieve higher recall. When integrated into graph-based indexes, vector quantization techniques such as PQ~\cite{jegou2010product} and SQ~\cite{aguerrebere2023similarity} can improve search efficiency with only a modest loss in accuracy. For collections beyond DRAM, DiskANN~\cite{jayaram2019diskann} stores the Vamana graph on SSD with a product-quantized in-memory representation, while PipeANN~\cite{guo2025lowlatency} further reduces query latency by overlapping computation with SSD reads. These indexes are widely used in industrial vector-search systems such as VSAG~\cite{zhong2025vsag} and Milvus~\cite{wang2021milvus}.

Several dynamic vector-search systems have been proposed to support real-time workloads. SPFresh~\cite{xu2023spfresh} adopts a clustering-based index but degrades in high-dimensional spaces. FreshDiskANN~\cite{singh2021freshdiskann} supports dynamic updates through lazy deletion and batch consolidation, but incurs deferred cleanup and expensive neighborhood repair. CleANN~\cite{zhang2026cleann} similarly defers physical cleanup while using query-adaptive consolidation and semi-lazy cleaning. IP-Vamana~\cite{xu2025place} uses graph search to recover candidate in-neighbors; Wolverine~\cite{liu2025wolverine} instead forms candidates from two-hop neighbors. Both still couple recovery with explicit cleanup of stale incoming edges. \merit's key insight is to repair the routing topology with surviving candidates, and use target versions to invalidate incoming edges outside the recovered set. This separation eliminates the need for exhaustive reverse-edge discovery during deletion.

\section{Conclusion}\label{sec:conclusion}

In this paper, we presented \merit, an efficient in-place deletion method for dynamic graph-based ANNS indexes. Bounded recovery and $k_r$-MST repair promote local connectivity along useful routes, while target versions invalidate all stale incoming edges missed during recovery. This division avoids explicit reverse-edge discovery and periodic full-graph maintenance. We analyze its deletion cost and the lifetime of the finite-width version counter. Experiments across multiple datasets demonstrate that \merit outperforms prior SOTA methods in update latency, post-update search stability, and scalability.

\bibliographystyle{ACM-Reference-Format}
\bibliography{add}

\end{document}